\documentclass[iop]{emulateapj}
\usepackage{graphicx}
\usepackage{color}

\newcommand{\eg}{e.g.}
\newcommand{\ie}{\textit{i.e.}}

\newcommand{\J}{\textsf{J}}

\newcommand{\Jz}{$J_z$}
\newcommand{\Hp}{H$+$}
\newcommand{\Hm}{H$-$}
\newcommand{\Sp}{S$+$}
\newcommand{\Sm}{S$-$}
\newcommand{\Ic}{$I_{\rm c}$}

\shorttitle{ELECTRIC CURRENTS IN FLARE RIBBONS}
\shortauthors{Janvier et al.}

\begin{document}

\title{ELECTRIC CURRENTS IN FLARE RIBBONS: OBSERVATIONS AND 3D STANDARD MODEL}

\author{M. Janvier
}
    \affil{Division of Mathematics, University of Dundee, Dundee DD1 4HN, Scotland, United Kingdom}
        \affil{LESIA, Observatoire de Paris, CNRS, UPMC, Univ. Paris Diderot, 5 place Jules Janssen, 92190, Meudon, France}
    \email{mjanvier@maths.dundee.ac.uk}
\author{G. Aulanier} \author{V. Bommier} \author{B. Schmieder} \author{P. D\'emoulin} \author{E. Pariat}
    \affil{LESIA, Observatoire de Paris, CNRS, UPMC, Univ. Paris Diderot, 5 place Jules Janssen, 92190, Meudon, France}


\begin{abstract}
{We present for the first time the evolution of the photospheric electric currents during an eruptive X-class flare, accurately predicted by the standard 3D flare model. 
We analyze this evolution for the February 15, 2011 flare using HMI/SDO magnetic observations and find that localized currents in \J-shaped ribbons increase to double their pre-flare intensity.
Our 3D flare model, developed with the OHM code, suggests that these current ribbons, which develop at the location of EUV brightenings seen with AIA imagery, are driven by the collapse of the flare's coronal current layer.
These findings of increased currents restricted in localized ribbons are consistent with the overall free energy decrease during a flare, and the shape of these ribbons also give an indication on how much twisted the erupting flux rope is.}
Finally, this study further enhances the close correspondence obtained between the theoretical predictions of the standard 3D model and flare observations indicating that the main key physical elements are incorporated in the model. 

\end{abstract}

\keywords{Sun: flares -- Sun: photosphere -- Sun: UV radiation -- Magnetohydrodynamics }

%

%
\section{Introduction}
\label{sect_intro}

Eruptive flares are amongst the main drivers of space weather as they inject solar material under the form of coronal mass ejections into the interplanetary medium \citep[\eg ][]{Bothmer1994,Forbes2006}.
These blasts of solar plasma are regularly at the origin of magnetic storms on Earth and other planets \citep[see][]{Gosling1991,Gonzales1991,Prange2004}.
Solar flares are classified in function of the peak flux of X-rays measured near Earth, and divided into different categories. The most energetic events are classified as X-class flares and have intensities starting from 10$^{-4}$~W/m$^2$.  The majority of X-class flares have been shown to originate from eruptive flares \citep{Wang2007}, making them important events to be understood. 

Eruptive flares have several observational characteristics, including the flare loops \citep[e.g.][]{Schmieder1995} and twisted magnetic field lines, identified as the flux rope at the origin of coronal mass ejections \citep[e.g.][]{Cheng11, Zhang2012, Patsourakos2013}.
Flare ribbons are also important flare observational characteristics, as they are the locations for the largest radiative emission increase of flares, in extreme ultraviolet (EUV) and in soft X-rays \citep[SXR, see e.g.][]{Schmieder1987, delZanna2006}. They are {a consequence of} the impact at the chromosphere of energetic particles launched from the coronal reconnection site \citep[e.g.][]{Reid2012}. During an eruptive flare, a two-ribbon structure is often observed \citep[e.g.][]{Asai2003,Ding2003}, with a typical \J-shape for both ribbons in the different magnetic polarities {\citep[see e.g.][]{Moore95,Chandra09,Warren2011}.}

Although the CSHKP 2D model \citep{Carmichael1964,Sturrock1966,Hirayama1974,Kopp1976} and its further extensions \citep[\eg ][]{Forbes1986,Shibata1995} explain some of the eruptive flare characteristics, it fails at rendering their intrinsic 3D nature, and therefore the complex evolution of {the flare loops, the} flux rope, and the \J-shaped ribbons. 

As such, different mechanisms for eruptive flares have been proposed \citep[see][ and references therein]{Forbes2006,Amari2003a,Amari2003b,Shibata2011,Aulanier2014}. One of these scenarios involves a torus-unstable flux rope \citep{Bateman78,Lin1998,Kliem2006}, and numerical simulations of such an evolution have been able to reproduce many of the eruptive flare characteristics \citep[\eg ][]{Aulanier2012,Janvier2013,Dudik2014}.
The expansion of the unstable flux rope leads to a feedback process as a thin current layer, formed underneath the flux rope, feeds the growing structure with twisted field lines via magnetic reconnection. This process is also at the origin of flare loops and their strong-to-weak shear evolution \citep[\eg ][]{Warren2011}.  This was quantified during the 2011, May 9 eruptive flare and was properly predicted in the present standard 3D model \citep{Aulanier2012}. 

The strong distortions of the magnetic field line mapping, referred to as Quasi-Separatrix Layers \citep[QSLs,][]{Demoulin1996a, Demoulin1997} are quantified by the squashing degree parameter $Q$ \citep{Titov02,Pariat2012}.  Strong current density regions develop at QSL locations \citep{Aulanier05b,Janvier2013}. The QSLs are therefore preferential locations for ideal MHD to break down, \ie\ for reconnection to occur. Indeed, both QSLs and current density structures show similar time evolution and shapes as flare ribbons (see Figures~3 and 15 in \citealt{Dudik2014} for the QSLs, and the outward motion of current ribbons shown in Figure 6 of \citealt{Aulanier2012}). 

	\begin{figure}
     	\centering
    	\includegraphics[bb=10 05 720 400,width=0.49\textwidth,clip]{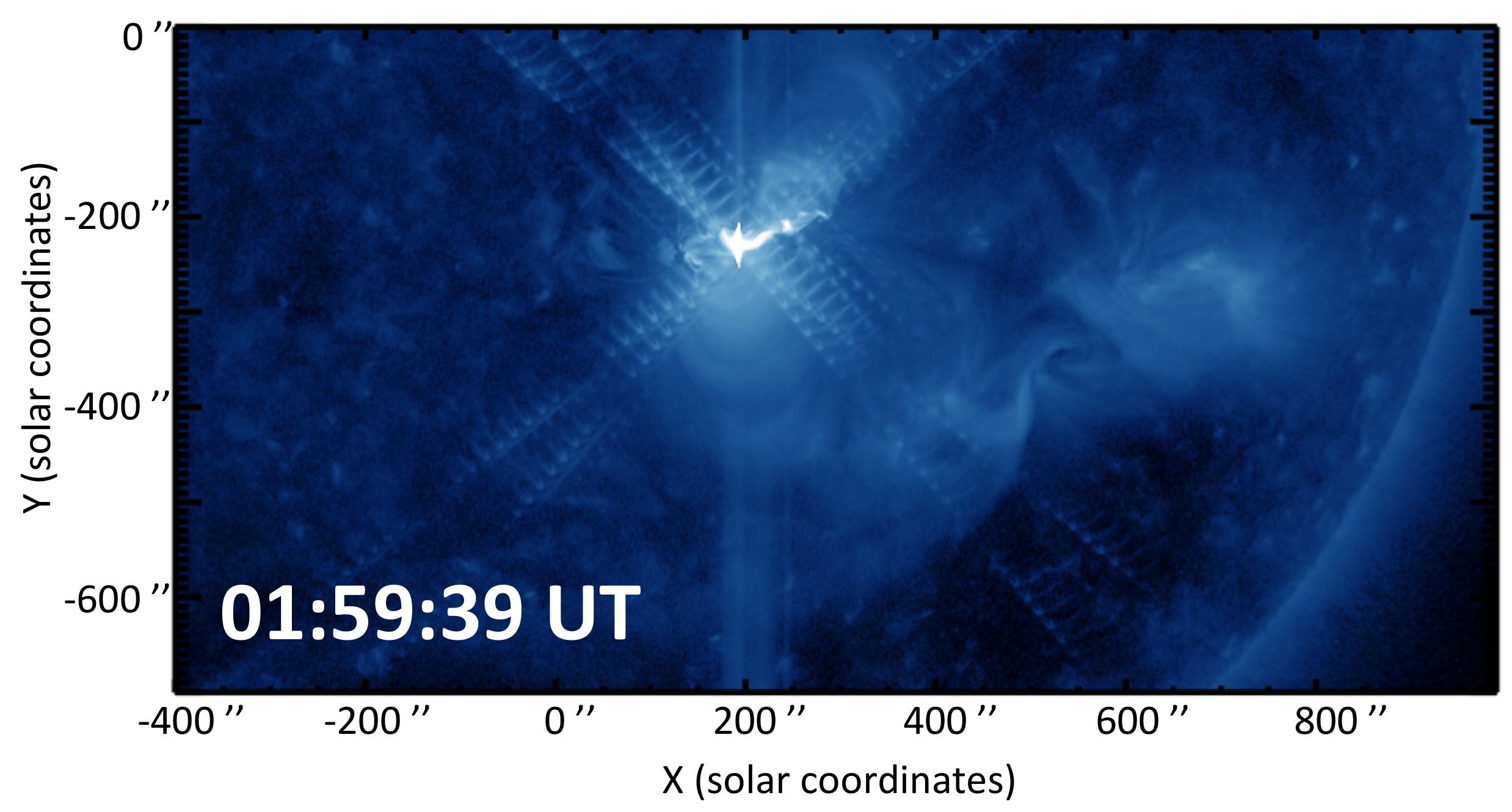}
	   	\caption{
AIA image in the 335~\AA\ filter showing the South-West quarter of the Sun during the peak of the 2011, February 15 X-class flare. This eruptive flare occurred in the central part of AR 11158, {close to the center disk. The center of the signal saturation and the diffraction pattern at the flare maximum marks the very localized flare region in spite of its large magnitude. }
				     	}
     	\label{figAIAoverview}
     	\end{figure}


As such, the present paper aims at quantifying the evolution of the photospheric electric currents and at directly comparing it with that predicted with the 3D numerical extensions to the standard flare model. More precisely, we study the vertical component of the electric current vector that we will simply denote as the vertical current in the following.
This is made possible by the high cadence and good spatial resolution of the HMI instrument onboard of the Solar Dynamics Observatory (SDO), and with the analysis of the vector magnetic fields at the photospheric level.


The remainder of the paper is as follows.
In Section~\ref{Sect:2}, we present a general overview of the February 15 flare and the SDO instruments, HMI and AIA. We also introduce the {UNNOFIT} inversion method that allows the calculation of the three components of the magnetic field, and therefore the photospheric vertical currents. In Section~\ref{Sect:3}, we study {quantitatively} the evolution of the photospheric current ribbons. These results are compared in Section~\ref{Sect:4} with the evolution of flare ribbons seen with the AIA instruments, as well as with the  current ribbons predicted in the 3D standard flare model developed in a numerical simulation. Finally, the implications of the present results are discussed in Section~\ref{Sect:5}, followed by a summary and a conclusion in Section~\ref{Sect:6}.
\section{Flare observations, magnetograms and current maps}
\label{Sect:2}
%

%
\subsection{Overview of SDO observations}
\label{Sect:2.1}
The \textit{Solar Dynamics Observatory} (SDO), launched in 2010, has a full solar disk field of view and comprises three instruments. Amongst them, the Helioseismic and Magnetic Imager \citep[HMI,][]{Scherrer2012, Schou2012} measures the vector magnetic fields in the Fe I absorption line at 6173~\AA . It has a spatial resolution of 1'' with a 0.5'' pixel size and a cadence of 45 s for the line-of-sight magnetic field. For the transverse magnetic field, the cadence is reduced to 12 min by averaging the four Stokes profiles before the inversion in order to enhance the signal over noise ratio.
This time cadence still allows us to obtain the magnetograms just before the impulsive phase and after the peak of the flare.

The Atmospheric Imaging Assembly \citep[AIA,][]{Lemen12,Boerner12} onboard SDO has 4 identical telescopes providing high temporal (cadence of 12~s) and spatial resolution (1.5'' with a 0.6'' pixel size) in different filters, 9 in the EUV wavelengths and one in visible wavelengths.
In Figure \ref{figAIAoverview}, a South-West portion of the Sun is shown during the impulsive phase of the flare. The wavelength of 335~\AA, corresponding to a temperature of 2.5$\times 10^6$K (see \citealt{ODwyer2010}), is taken throughout the paper since the flare ribbons are the least saturated in this filter (as it allows us to see coronal structures with high temperature). 
		\begin{figure}
     	\centering
     \IfFileExists{fastCompil.txt}{
    	\includegraphics[width=0.48\textwidth,clip]{fig_2.png}
	                              }{
    	\includegraphics[bb=100 160 500 800,width=0.5\textwidth,clip]{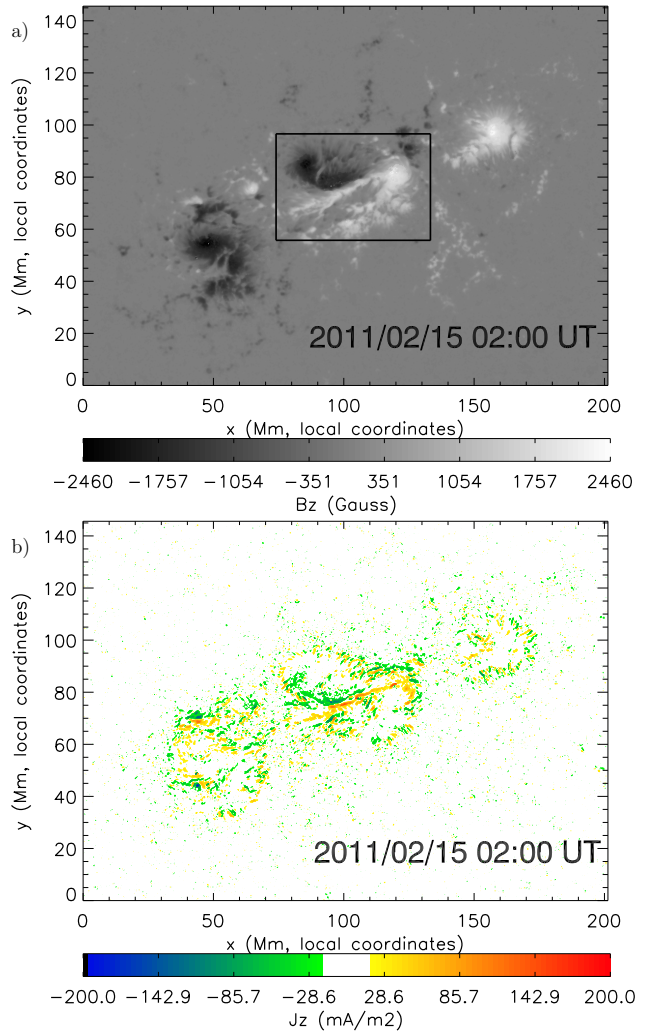}	                         
 	                              }
	   	\caption{{({\bf a}) Vertical component magnetogram} of the AR 11158 at 02:00~UT, where the central two polarities (in black box) are the one studied throughout the paper. ({\bf b}) Maps of {the vertical current component} \Jz\ at 02:00~UT.  {The white region of the color scale is selected with  the noise level defined as $|$\Jz$|$=20 mA.m$^{-2}$.} The two maps are shown in solar local coordinates.
		     	}
     	\label{figBzJz}
     	\end{figure}

\subsection{{Overview of the studied flare} }
\label{Sect:2.2_Overview}

{In order to analyze photospheric structures, observations near the central region of the solar disk is required, so as to avoid the degradation of observations due to projection effects. The February 15, 2011 event, occurring in active region 11158, is a very good candidate, for the strength of the event (X2.2) and for its location near the center disk (Figure~\ref{figAIAoverview}). It is also a good candidate because such a compact active region flare allows better measurements of the horizontal magnetic field vector in the related sunspots, the magnetic field being more intense in those regions.}

		\begin{figure*}
     	\centering
     \IfFileExists{fastCompil.txt}{
    	\includegraphics[width=0.8\textwidth,clip]{fig_3.png}
	                              }{
    	\includegraphics[bb=70 80 1150 1505,width=0.75\textwidth,clip]{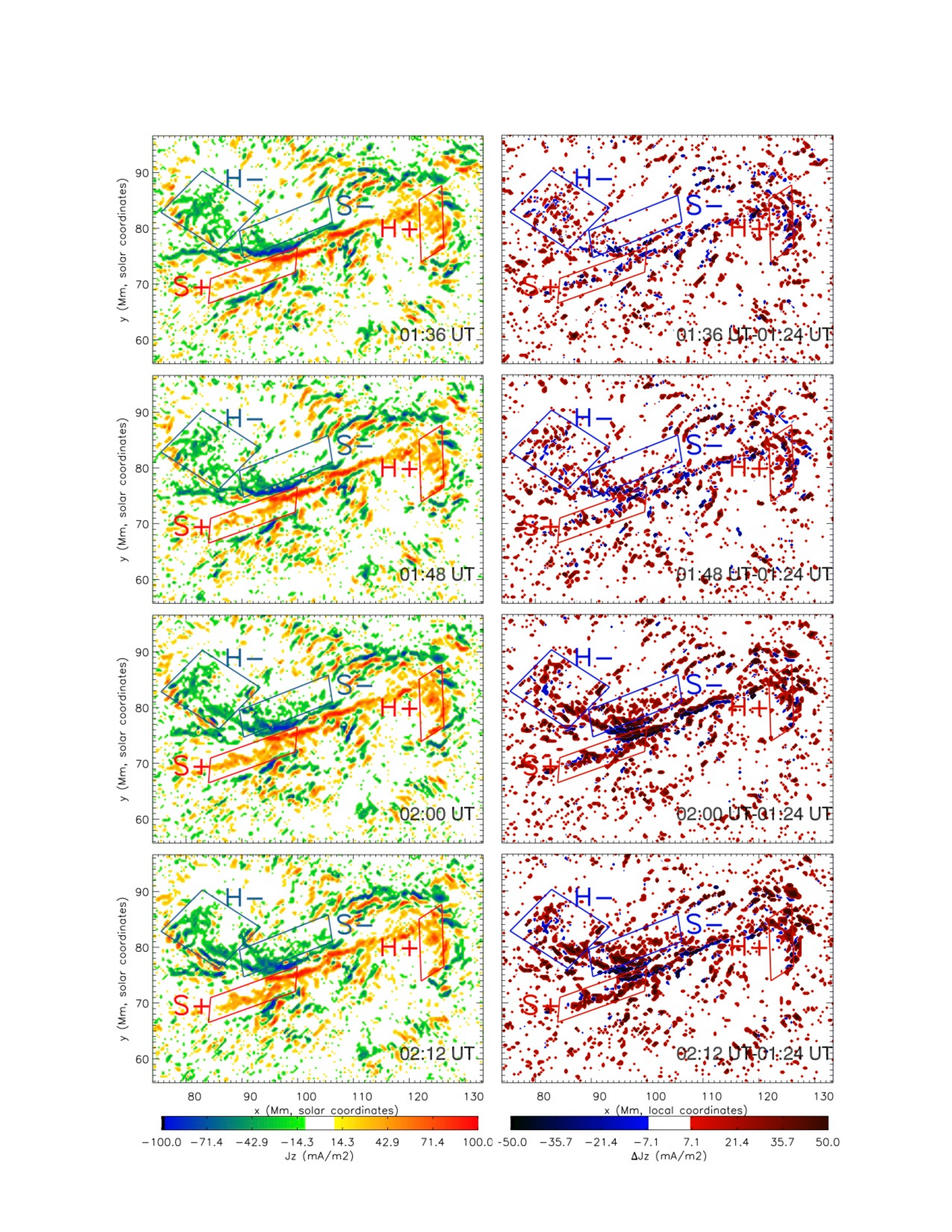}	                         
	                              }
	   	\caption{
   {\textbf{Left}: maps of the vertical current density} \Jz\ in the central bipole (see black square of Figure \ref{figBzJz}a) at two times before and after the impulsive phase of the flare. 
The white interval corresponds to removed \Jz\ signal at $\pm$ the noise level. 
\textbf{Right}: \Jz\ base-difference of the direct currents ($\vec{J}_z (t) \cdot \vec{B}_z(t)>0$) with the base taken at 01:24~UT, \ie\ well before the flare starts. The different squares show the hook and the straight parts of the \J-shape current ribbons where changes are the most prominent. {They are further studied in Figure~\ref{figJint}.}
		     	}
     	\label{figJzzooms}
     	\end{figure*}

The active region 11158 is part of a complex region involving four magnetic polarities (Figure~\ref{figBzJz}) that evolved over several days, leading to 2 M-class flares (February 13 \& 14) and one X-class flare on February 15 \citep{Schrijver2011,Gosain2012,Inoue2013}. The February 13 2011 M6.6 flare was investigated in details by \citet{Liu2013}, who also showed the formation of two flare ribbons and observational evidences of magnetic reconnection via the tether-cutting reconnection model. Although the magnetogram shows a quadrupolar structure for the magnetic polarities, the X2.2 flare occurred in the central bipole region, as the flare ribbons extending from $[x,y]=[170'',-240'']$ to $[240'',-210'']$, so over a very limited portion of the Sun (Figure~\ref{figAIAoverview}). 
{\citet{Sun2012}, \citet{Wang2012} and \citet{Petrie2013} found that the horizontal photospheric magnetic field was enhanced and more sheared in between the flare ribbons by about 30\% after the X2.2 flare, while the vertical field component had no significant modifications. Note that \citet{Petrie2013} also showed maps of the vertical current density, but without any detailed analysis as we perform below.
Non-linear force free field extrapolations complemented the vector magnetograms by allowing the study of the coronal magnetic field evolution \citep{Sun2012,Inoue2013,Liu2013,Dalmasse2013}. In particular, the topology of this active region, along with the presence of a flux rope and QSLs, was investigated by \citet{Zhao2014}.}

	\begin{figure*}
     	\centering
    	\includegraphics[bb=-10 -5 2060 700,width=1\textwidth,clip]{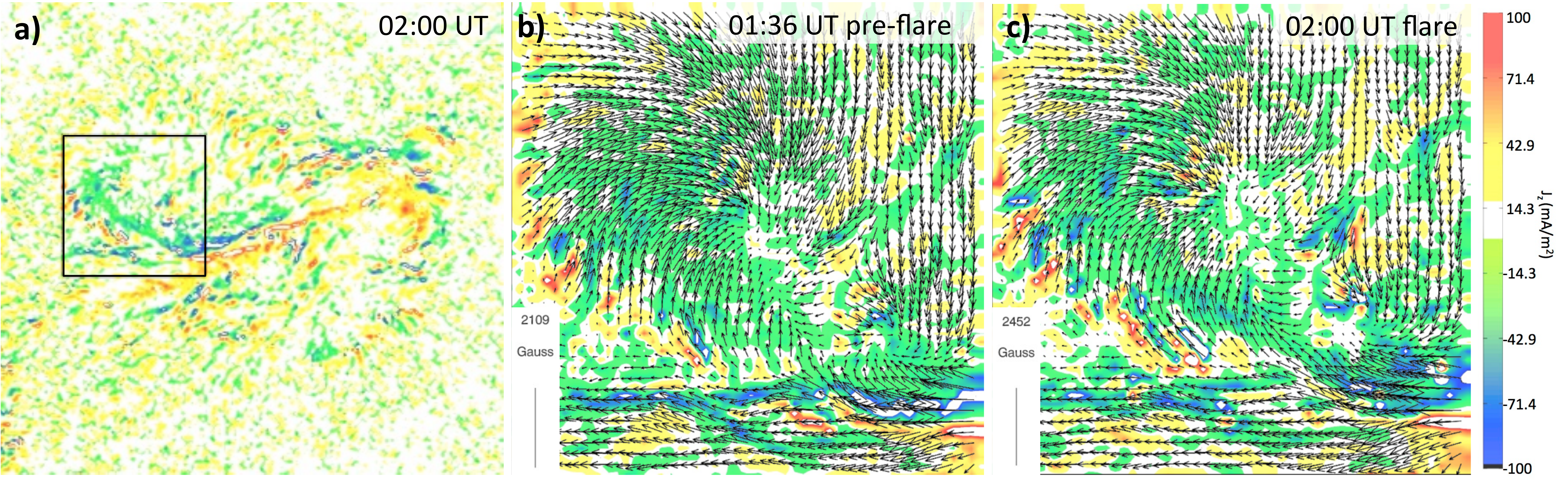}
	   	\caption{Vertical current density \Jz\ and horizontal magnetic field vector maps. The overview of the \Jz-map is shown in panel ({\bf a}) where the square corresponding to the zooms in panels ({\bf b}) and ({\bf c}) is indicated. The zoom is on the negative current ribbon hook region and the vector magnetic field vectors are evolving from before ({\bf b}) to the end of the impulsive phase ({\bf c}). These changes of the magnetic field are coherent around a large area around the current and flare ribbons (shown in Figure~\ref{figJcomparison}a,d). 
		     	}
     	\label{figBvector}
     	\end{figure*}

\subsection{UNNOFIT inversion method {for vector magnetograms}}
\label{Sect:2.3_UNNOFIT}


We downloaded the HMI level-1b IQUV data, and inverted them with the Milne-Eddington inversion code UNNOFIT \citep[see][]{Bommier2007}. We selected a region covering AR 11158, and applied a solar rotation compensation to select the same region over 4 hours of observation. We thus treated 20 maps, from 2011/02/15 00:00 UT to 2011/02/15 03:48 UT.

The specificity of UNNOFIT is that a magnetic filling factor is introduced to take into account the unresolved magnetic structures, as a free parameter of the Levenberg-Marquardt algorithm that fits the observed set of profiles with a theoretical one. However, for further application we used only the averaged field, i.e. the product of the field with the magnetic filling factor, as recommended by \citet{Bommier2007}. The interest of the method lies in a better determination of the field inclination. We tried to compensate the rather poor spectral HMI resolution by interpolating with a cubic spline function between the 6 spectral points. However, we found that the inversion results were better, with an apparent better spatial resolution, without the interpolation, i.e. with 6 spectral points adjusted on the computed theoretical profile of the 4 Stokes profiles. In the code, the loop on the pixels is parallelized by using the OpenMP method. We had 12 cores running in parallel for the inversion of one map on our computer, and since we were allowed to invert five maps simultaneously, it took one hour of computation to treat one hour of observational data.

After the inversion, the $180^{\circ}$ remaining azimuth ambiguity was resolved by applying the ME0 code developed by Metcalf, Leka, Barnes and Crouch \citep{Leka2009} and available at http://www.cora.nwra.com/AMBIG/.  After resolving the ambiguity, the magnetic field vectors were rotated into the local reference frame, where the local vertical axis is the $Oz$ axis.

\subsection{Magnetogram evolution}
\label{Sect:2.4_Magnetograms}

{Figure \ref{figBzJz}a} shows the four magnetic polarity regions almost located on the central part of the solar disk. The coordinates are given in the local solar reference (\ie, not in the line-of-sight coordinates). The four polarities started emerging at the end of February 10, 2011, and as they kept evolving, the trailing negative polarity of the west bipolar system moved closer to the positive leading polarity of the east bipolar system. After several flares involving the magnetic reconnection between the two bipolar fields, the two central polarities are strongly linked by magnetic arcades.  These two polarities encounter each other with a close ``fly-by'', leading to a strong shear of the magnetic arcades {\citep[see][]{Toriumi2013}.} This provides the conditions for further flares to occur, including the studied X2.2 flare, in between these two polarities.

The X-class flare of February 15 took place in the central bipolar region (contoured with a black square in Figure \ref{figBzJz}a). As such, although the whole domain presents a rather complex quadrupolar region, the X-class eruptive flare can be studied only by focusing on the central region. 
Figure \ref{figBzJz}a shows the polarities of $B_z$ after the peak of the flare, which are not significantly different from before the flare, as shown by \citet{Sun2012} and \citet{Petrie2013}.

\subsection{Mapping of the photospheric currents}
\label{Sect:2.5_Mapping}


Figure \ref{figBzJz}b shows the vertical component \Jz\ of the photospheric current at 02:00~UT of AR 11158. The current density \Jz\ is directly derived from the magnetic field components via a centered difference method from the equation: $\nabla \times \vec{B} = \mu_0 \vec{J}$ written in the local frame.
Since the vertical component of the magnetic field does not change much in time, we are interested in the changes of the vertical current density changes during the eruptive flare, that are therefore directly related with the changes of the horizontal magnetic field. This vertical current is the only one which could be computed from a vector magnetogram observed with one spectral line.  It is still the most interesting component for flare studies as those currents are expected to enter in the corona. Knowing the vertical component of the electric current density is also of importance for extrapolating the coronal magnetic field \citep[\eg ][]{Wheatland2013}.

The \Jz-map grid has a mesh size {of $\approx 370$ km.} The map is centered on the central bipole and the coordinates are the local solar coordinates.
{The values of $|J_{z}|$ below the noise level of {20 mA.m$^{-2}$} are not shown.}
This level is estimated by the formula {$\delta B_T/(\mu_0 \Delta x$) \citep[see][]{Gary1995} } where $\delta B_T$ is the transverse field error, and could be of the order of $\sim 100$~G  \citep[see \eg ][]{Wiegelmann2010} and $\Delta x = 0.5''$ is the grid size.

In the region of interest (the two central polarities, see the square in Figure \ref{figBzJz}a), we can already point out the strong current density regions near the polarity inversion line (PIL). The negative (resp. positive) density currents (in green, resp. in yellow/red) mostly appear in the negative (resp. positive) polarity. However, mixed densities also appear further away from the PIL, probably due to the presence of the interlocking-comb structure in the penumbra of the sunspots \citep{Thomas1992, Solanki1993}, as already seen in other current maps from \citet{Venkatakrishnan2009}.

Near the PIL, the positive and negative current density regions align with each other, suggestive of opposite current ribbons. 
Those current densities extend upward/downward while contouring the negative/positive polarities, {so that they both have a \J-shape.} 
The time evolution of those high current density regions is investigated in {Section}~\ref{Sect:3}.

\section{Evolution of the current density ribbons during the flare}
\label{Sect:3}
%

%
\subsection{Evolution of the electric current ribbon shapes}
\label{Sect:3.1}

We show in Figure \ref{figJzzooms} a zoom of the photospheric current maps in the central bipole at four different times before (01:36~UT and 01:48~UT) and after (02:00~UT and 02:12~UT) the impulsive phase of the flare. This zoom corresponds to the black square indicated in Figure \ref{figBzJz}a, and the times are restricted by the temporal resolution of HMI (12 min). The strong and coherent currents are dominantly direct currents along the magnetic field ($\vec{J} \cdot \vec{B}>0$ with a dominant positive current helicity) as found by \citet{Petrie2013}.
We also show 4 squares indicating different regions, such as the hook and the straight part of the \J-shaped current ribbons.
 This allows the reader to quickly pinpoint the different changes occurring during the flare evolution. We refer to \Hm\ (resp. \Hp) for the hook part of the ribbon in the negative (resp. positive) polarity, and to \Sm\ (resp. \Sp) for the straight part of the ribbon in the negative (resp. positive) polarity.

The left column {of Figure~\ref{figJzzooms}} shows the different times of the zoomed \Jz-maps while the right column shows the maps of the base-difference of the direct electric current density. The base reference image is chosen at 01:24~UT, quite before the onset of the flare. 
We only show the evolution of the direct currents which stay direct (\ie\ we show the currents both satisfying  {$J_z^{\rm phot}$(01:24) $B_z^{\rm phot}$(01:24)$>0$} and $J_z^{\rm phot} (t)  B_z^{\rm phot}(t)>0$).

\begin{figure*}
     	\centering
   	\includegraphics[bb=10 210 600 600,width=1\textwidth,clip]{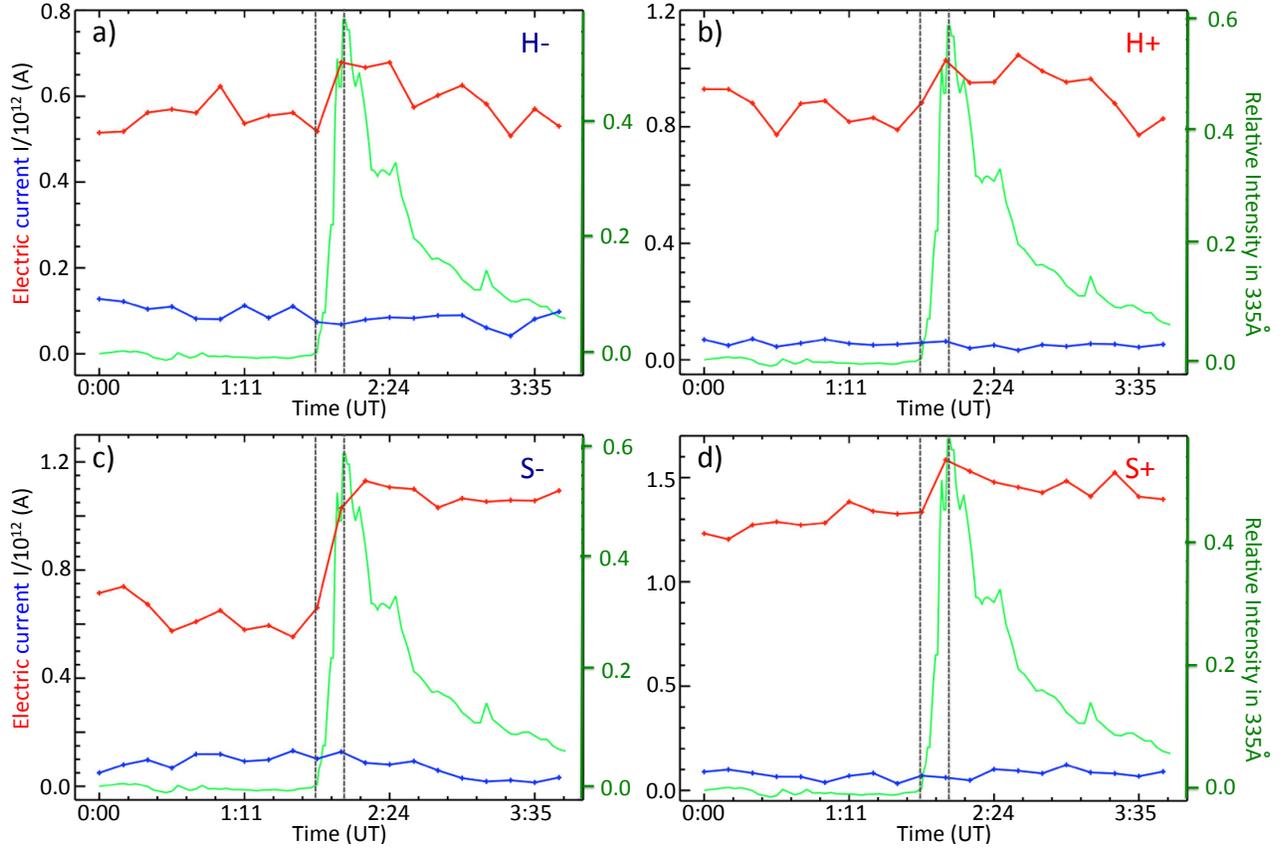}
		   	\caption{Evolution of the direct electric current (in red) and the return electric current (in blue) before, during and after the X-class flare in the \Hm\ ({\bf a}), \Hp\ ({\bf b}), \Sm\ ({\bf c}), \Sp\ ({\bf d}) regions (as defined in Figure~\ref{figJzzooms}). The light curve for the full Sun signal in the AIA 335~\AA\ channel is also plotted on the same graphs in green. This light curve is the relative intensity normalized to the intensity at 0:00~UT.
The dotted vertical lines indicate the beginning and the end of the flare impulsive phase.
				     	}
     	\label{figJint}
     	\end{figure*}

There is almost no evolution between 01:36~UT and 01:48~UT {(Figure~\ref{figJzzooms}).} 
However, most of the changes can be seen between 01:48~UT (before the flare peak) and 02:00~UT (after the flare peak).  In \Hm, the hook shape becomes structured at 02:00~UT, while the signal is {more randomly oriented} before the flare peak. There is a signal increase at the base of the hook, around $x=85$~Mm and $y=78$~Mm, as it is clearly seen in the base-difference (right) figure at 02:00~UT. The hook also becomes more consistent in the \Hp\ region with an increase of the current density around $x=123$~Mm, $y=82$~Mm. There is also a strong increase of the current density in the straight-parts of the ribbons (see the dark red ribbons in the base-difference image). In the \Sm\ region, there is a broadening of the current ribbon, as \Jz\ increases in the higher part ($y \approx 80$~Mm) of the region. The positive current ribbon does not broaden as much, but elongates toward the lower $x$ region {with an} increase in the current density.  All these changes remain at 02:12~UT when the flare brightness decreases significantly (see the light curve in 335 \AA\ in the later Figure~\ref{figJint}), indicating that the signals we analyze are unlikely to be related to polarization artifacts.

\subsection{Arguments against ribbon artifacts from polarization effects}
\label{Sect:3.2}

Although the current structures that we see appearing after the flare at 02:00~UT {have a coherent spatial organization,} polarization artifacts should be seriously considered especially since these appear at flare ribbon locations. Indeed, during solar flares, such effects have often been reported {at the chromospheric level \citep[see][and references therein]{Henoux2013}. The} non-thermal particle beams and neutralizing return currents are generally cited as possible sources for {an extra polarization signal.} However, more recently, \citet{Stepan2013} showed that this polarization is not due to energetic particles impacting the chromosphere, but to a radiation anisotropy near the chromospheric ribbons.
Since this polarization effect is mainly seen for chromospheric lines (e.g. H$\alpha$, Na D2) in which ribbons are always very bright, the present photospheric data, which do not reveal any obvious signal in the intensity (Stokes I, in the Fe I 6173~\AA\ line), should avoid such artifacts. 
However, since the interest of this paper is to analyze in details the evolution of the high current density region, we {check in the following} whether current signals are due to coherent, physical changes of the magnetic field.

So as to verify the reality of the current ribbons shown above, we show in Figures~\ref{figBvector}b,c, two zooms of the same region, the hook of the \J-shape negative current ribbon (\Hm), also indicated with the square in Figure \ref{figBvector}a. These two zooms also show the vector map of the horizontal magnetic field superposed on top of the \Jz-maps, and represent two times taken before (b) and after (c) the flare peak. 

The horizontal field vector $\vec{B}_{h}$ has well resolved vectors that smoothly rotate across the current ribbon, all along the straight and curved part of the hook shown in Figure \ref{figBvector}.
This is a clear indication of a consistent magnetic field signal free of artifacts, as otherwise the $\vec{B}_{h}$ directions would appear {more randomly oriented}, especially near the edges of the ribbons during the flare (Figure \ref{figBvector}c). Also, the rotation of the vector field is a very good indicator of the presence of a current layer. Indeed, the sudden rotation of $\vec{B}_h$ should be expected in the present of a high current density region, as is depicted in the Harris sheet model in 2.5D with a guide field along the current and field component across the sheet {\citep{Harris62}}.

 Interestingly, these two figures also show the evolution of the horizontal component of the magnetic field during the flare. In particular, the changes of the vector field characteristics are readily seen in the coherent ribbon structure that is formed at the base of the hook: both the direction and intensity of the vectors are changing coherently. These results agree with the quantification of irreversible changes during the flare, in the photospheric horizontal magnetic field quantified in \citet{Sun2012} and \citet{Wang2012}. We conclude that the electric density changes are not mere artifacts due to polarization effects.

		\begin{figure*}
     	\centering
     \IfFileExists{fastCompil.txt}{
    	\includegraphics[width=\textwidth,clip]{fig_6.png}
	                              }{
    	\includegraphics[bb=10 10 2050 1000,width=1\textwidth,clip]{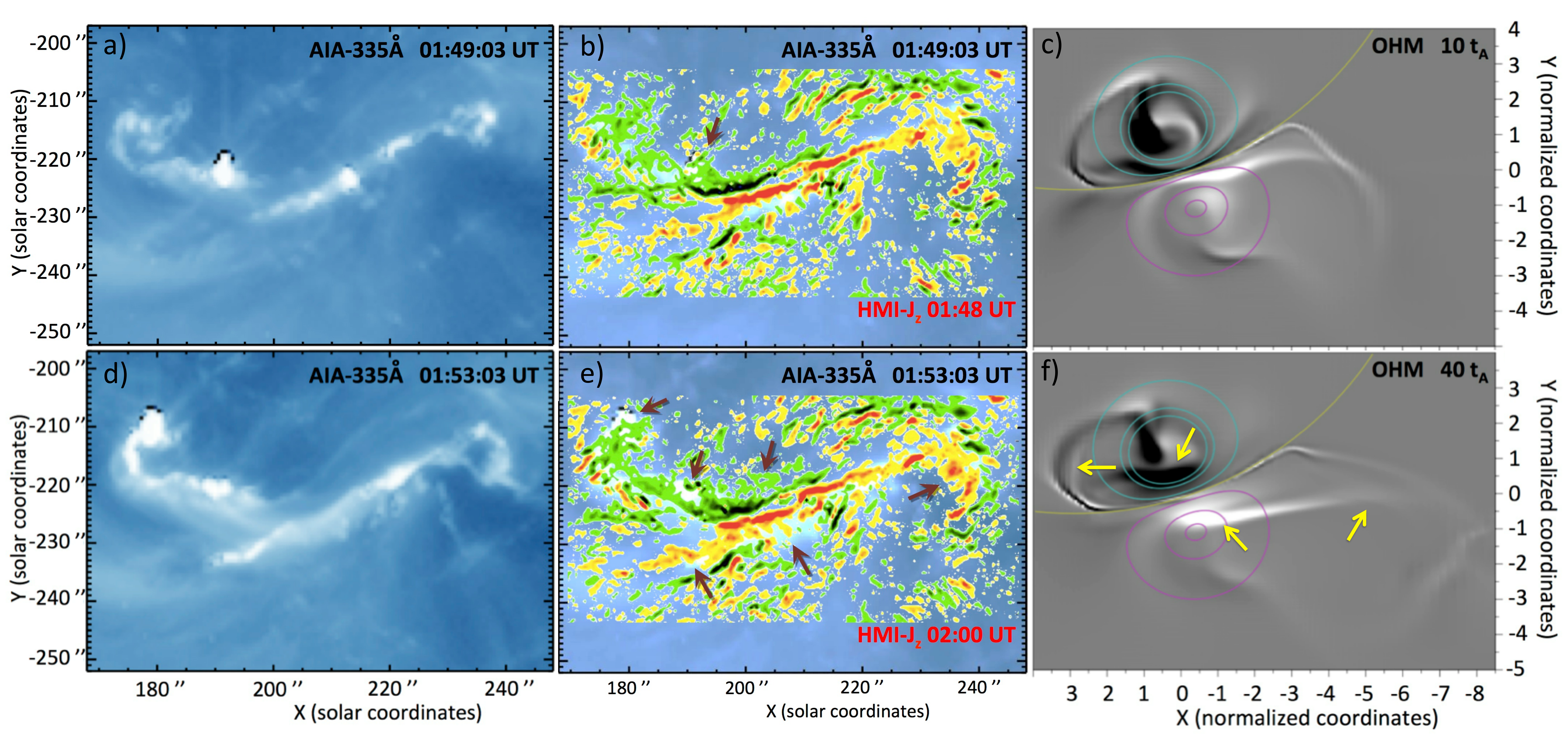}	                         
	                              }
		   	\caption{\textbf{Left}: Flare ribbons before ({\bf a}) and after ({\bf d}) the peak of the flare as seen in the SDO/AIA 335~\AA\ filter. \textbf{Middle}: {Vertical current density maps, or \Jz\ maps,} from SDO/HMI before ({\bf b}) and after ({\bf e}) the peak of the flare superposed over the saturated AIA 335~\AA\ filtered images of the ribbons. {Note that since flare ribbons are very well co-aligned with the current ribbons, they are almost fully covered by the superposed high \Jz\ regions.} The brown arrows show similar structures found for \Jz\ and the ribbon maps and point out the differences after the flare. \textbf{Right}: \Jz\ photospheric maps from OHM 3D simulations of an eruptive flare. The yellow arrows point out the differences seen around the end of the simulation compared to the beginning.
					     	}
     	\label{figJcomparison}
     	\end{figure*}

%
\subsection{Current signal evolution}
\label{Sect:3.3}

Since the current density significantly changes before and after the peak of the flare in the different regions \Hp, \Hm, \Sp, \Sm, we study in the following the evolution of the electric current intensity \Ic\ in these different regions. \Ic\ is readily calculated by the integration of the current density over the different surfaces considered (\Ic= $\mathop{\int \!\!\! \int} J_z dS$ where $S$ is the surface comprised in the four regions of interest). 

We report in Figure \ref{figJint} the evolution of the electric current integrated from the direct current densities ($J_z^{\rm phot} (t) B_z^{\rm phot}(t)>0$, shown in red) and the return current densities ($J_z^{\rm phot} (t) B_z^{\rm phot}(t)<0$, shown in blue). We have also added the light curve 335~\AA\ AIA filter (green curve), showing the relative light intensity calculated as $I_{\rm rel}=I_{335 {\rm \AA}}(t)/I_{335 {\rm \AA}}(0~{\rm UT})-1$.

In all the different regions, the signal for the return currents (\ie\ $\vec{J} \cdot \vec{B} <0$) is significantly low compared to that of the direct currents. Furthermore, there is no perceptible evolution of these return currents before, during and after the flare peak. On the contrary, the direct electric currents (in red) are significantly changing during the impulsive phase of the flare. Indeed, all panels show an increase in the current intensity in the same time range as the light emission increase within the limits of the 12 min temporal resolution of HMI and of the fluctuations before and after the flare. 
Then, \Ic(direct) increases by a factor of 1.4 {in \Hm\ during the flare and} by about 1.3 in \Hp. In the regions \Sm\ and \Sp, \Ic(direct) increases by 2 and 1.2 respectively. 
Moreover, during the decaying phase of the flare, the currents evolve slowly (Figure~\ref{figJint}), when the flare ribbons have strongly weakened in intensity.

This is the first quantification of the electric current evolution in current ribbons during an eruptive flare. In summary, we found three main characteristics: first,
the evolution of the coherent electric ribbons (Figure \ref{figJzzooms}); second, the decrease of the curvature radius of the horizontal magnetic field $\vec{B}_h$ across the current ribbons (Figure \ref{figBvector}); and finally, the slow evolution of the electric current after the sudden increase during the flare impulsive phase (Figure \ref{figJint}).  All these further confirm that the current evolution observed is not due to a modification of the light polarization due to energetic particles or enhanced radiation.

Predictions of the evolution of the flare, and especially the flare loops and the flux rope, have been given in a series of papers (see \citealt{Aulanier2012, Janvier2013}). In the following section, we propose to look at the results obtained via a 3D simulation of an eruptive flare in light of the present observations of the flare and electric ribbons.

%
\section{Currents in the 3D standard model and relation with the flare ribbons}
\label{Sect:4}
		\begin{figure}
     	\centering
    	\includegraphics[bb=0 0 3000 1750,width=1\textwidth,clip]{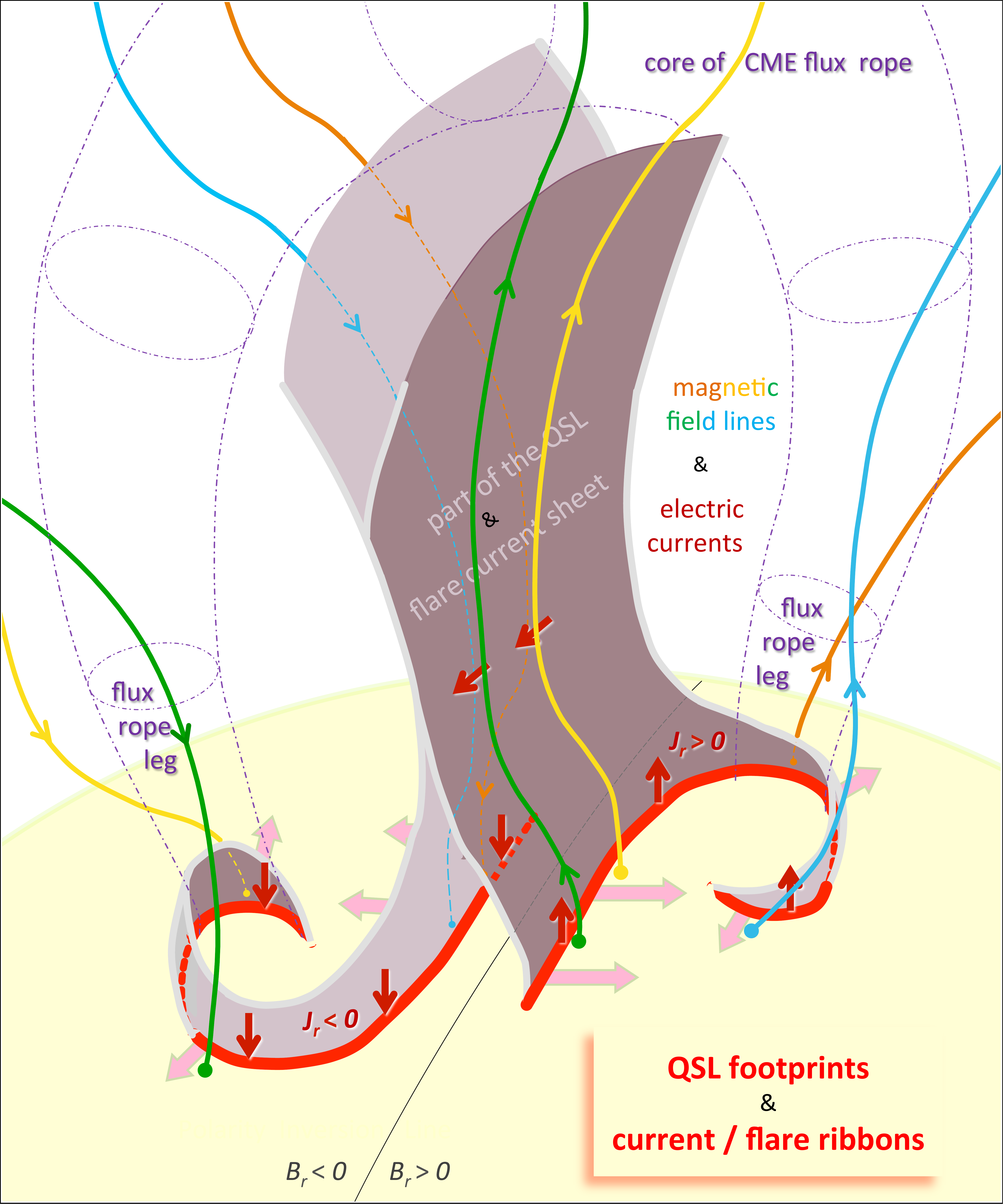}	                      
		   	\caption{{Cartoon for the standard model in 3D where four pre-reconnected magnetic field lines are shown. The grey area represents parts of the 3D volume of the QSL and the current layer. The 3D volume of the QSL has been restricted to only show the location of strong currents, and limited to a low height (below the flux rope). For a complete structure of the QSL volume, one can refer to the paper by \citet[Figure7]{Titov2007}. 
			Note that this cartoon remains valid for magnetic polarities of opposite signs as presently displayed, and/or for a configuration mirroring the present one, i.e. with two inverted-\J\ instead of two foward-\J. With one of these changes only, all the directions for the electric current vectors have to be reversed.
With both changes together, these directions remain the same. }
							     	}
     	\label{figmodel}
     	\end{figure}

%
\subsection{Flare ribbons evolution}
\label{Sect:4.1}

The evolution of the flare ribbons during the February 15, 2011 event are analyzed with the SDO/AIA instrument. We choose the 335~\AA\ filter since it gives a good compromise between the saturation level, the possibility to clearly see the whole ribbons and for a rather long period (\ie, the flare loops saturate in 335~\AA\ later after the formation of the ribbons). The comparison with other filters can be seen in Movie 3 of \citet{Schrijver2011}. We show in Figure \ref{figJcomparison}a, d, the flare ribbons before and during the impulsive phase of the flare. 
Similar changes as pointed out for the evolution of the current density ribbons can be readily seen: the hooks for both flare ribbons become more consistent at 01:53:03~UT, while the emission intensity increases. The hooks also expand and then become rounder (see for example the flare ribbon of the positive polarity, $(x,y) \approx (240'',-210'')$, {where the hook surrounds a dimming region indicating the presence of a flux rope}). During the flare peak, the straight parts of the ribbons also broaden (the ribbon in the negative polarity extends toward higher $y$ values, and inversely for the ribbon in the positive polarity). Moreover, the ribbon in the positive polarity also elongates (the end of the straight part extends from $x \approx 200''$ to $190''$). 

All those different changes are comparable to that of the current density ribbons as studied in Section~\ref{Sect:3}.  {Figures~\ref{figJcomparison}b,e} show the two \Jz-maps obtained from HMI (similarly to Figure \ref{figJzzooms} but in the observing frame) superposed on top of the AIA co-aligned images. The brown arrows point the different changes observed in both the \Jz\ ribbons and the flare ribbons. These two figures demonstrate the similar nature of the current ribbons and the flare emission ribbons.

%
\subsection{Current evolution in the 3D standard flare model}
\label{Sect:4.2}

The evolution of the current density can also be studied by means of a numerical simulation. {In the following, we present this evolution resulting from a 3D MHD simulation of an eruptive flare. This $\beta=0$ simulation performed with the OHM code reproduces the evolution of an initially torus-unstable flux rope} \citep{Aulanier2010,Aulanier2012}. The setups of these simulations were inspired from those of the pioneering 3D simulations from \citet{Amari2003a,Amari2003b} of the flux cancellation model. As the flux rope expands, regions of high current densities are created, where magnetic reconnection takes place.  The evolution of reconnected field lines, the topology of the magnetic field (including QSLs), as well as the 3D intrinsic reconnection process have been analyzed in details in \citet{Aulanier2012} and \citet{Janvier2013}. Especially, Figure~1 in \citet{Janvier2013} shows the formation of a coronal current layer and its thinning process as the flux rope continues to expand. It also shows the direct connection with the QSLs, and the broadening of the cusp underneath the coronal current structure can be directly linked with the footprints of the photospheric currents.

In Figure \ref{figmodel}, we have represented the main features of this 3D model. The grey areas represent the coronal current layer and its extension to the photosphere. More accurately, we only represent here the locations of high electric current densities. These locations are also locations of QSLs, where the magnetic field connectivities are the mostly distorted (see Section \ref{sect_intro}).The footprints of QSLs and current layer are \J-shaped, as can also be seen in images from the simulation in Figures~\ref{figJcomparison}{c,f}\footnote{In the present paper, the signs of the magnetic field vector $\vec{B}$, and therefore of \Jz, have been inversed compared to \citet{Janvier2013}. This allows a closer correspondence with the magnetogram for the Feb. 15, 2011 event.}. The hooks surround the legs of the flux rope, and four magnetic field lines have been represented, prior to reconnection.

The photospheric currents ribbons separate as time goes by (see pink arrows in Figure \ref{figmodel}), similarly to the flare ribbons. Indeed, Figure 6 in \citet{Aulanier2012} shows that reconnected field lines, so flare loops, are anchored in high photospheric current density regions. As flare loops are known to be rooted close-by to flare ribbons, the evolution of the flare ribbons is expected to trace the evolution of current ribbons, as observed.

The photospheric currents are shown in Figures~\ref{figJcomparison}{c,f} in the early and later stages of the numerical simulation. 
The contour-plots in cyan and pink trace respectively the negative and the positive vertical magnetic field.  Both the magnetic polarities and the current ribbon positions are comparable to the present flare observations, as follows.
The outward motion of the current ribbons, as expected in the flare ribbons evolution, is reproduced. 
These are pointed out with the yellow arrows.
{These shapes and displacements of the flare ribbons were also recovered by \citet{Inoue2014} with MHD simulations starting from an NLFFF extrapolation of the same active region, as follows (see their Figure 7 and Figure \ref{figJcomparison} of this paper)}.

These shapes and displacements of the flare ribbons were also
 recovered by Inoue et al. (2014) with MHD simulations starting
 from an NLFFF extrapolation of the same active region (compare
 their Figure 7 with our Figure 6)

Furthermore, the current ribbons elongate with an increase of their current density in the straight parts, and the hooks expand while becoming rounder. This is due to the building process of the flux rope: as magnetic reconnection takes place, part of the reconnected magnetic field lines further build up the envelope of the flux rope. This is also shown in Figure 17 of \citet{Dudik2014}, where the hooks of the current ribbons surround the newly reconnected magnetic field lines and therefore expand as the flux rope becomes bigger. The 3D structure of the current layer is also shown for a numerical simulation in Figure 11 of \citet{Kliem2013}, and especially the hook part that contour the flux rope legs.

We also point out the existence of return currents contouring the hooks of the electric current ribbons: the positive current hook in white in the positive polarity is contoured with black (\ie\ return current), while the black hook in the negative {polarity, indicating direct negative current,} is surrounded by white signatures of return currents. This feature also appears in the observations of the photospheric currents (Figures \ref{figJcomparison}{d,e}) where both direct current hooks (yellow and green in respectively the positive and negative polarities) are surrounded by opposite colour patches.
{We conclude that} all the observed characteristics present in the $J_{z}$ maps (Section~\ref{Sect:3.3}) are predicted by the 3D numerical model.

%
\subsection{{Main model characteristics needed to understand flare observations}}
\label{Sect:4.3}

The fact that similar changes are found for the evolution of the current ribbons in the simulation is quite remarkable since the simulation is not a data driven simulation but on the contrary, has theoretical boundary conditions solely designed to model a non-symmetric bipolar magnetic configuration as typically observed in mostly bipolar active regions.   The boundary driving is also generic of the shear concentration around PILs \citep{Schmieder1996} and of the progressive dispersion observed in active regions.   Such driving was only used to build the stressed magnetic configuration which contains mostly a magnetic sheared arcade with a moderately twisted flux rope inside.   The photospheric driving was stopped during the eruption, which is driven only by the magnetic forces growing after the torus instability/loss of equilibrium occurred.  Magnetic reconnection is occurring during all the evolution mostly at separatrices and QSLs.

To conclude, the 3D MHD simulation only incorporated a simple bipolar magnetic configuration with few generic processes typically observed in the evolution of active regions. The simulations for the torus instability conditions of a flux rope were also performed and the results published {\citep{Aulanier2010}} well before the presently studied flare occurred on the Sun. This is a clear indication that the key ingredients for an eruption are incorporated in the MHD simulation.

The results presented above are also striking considering that at the photospheric level, there is a finite plasma $\beta$ while the MHD simulation is realized in the zero $\beta$ limit. However, the electric currents shown in {Figure~\ref{figJcomparison}c,f} are only the traces at the lower boundary from electric currents present in the coronal volume.   They are present along a thin complex volume which follows the QSLs system.   The photospheric currents therefore have a coronal origin, and they are simply mapped down to the lower boundary along the coronal field lines. 

The flare ribbons shown in {Figure~\ref{figJcomparison}a,d}, are formed at the bottom of the corona and they are directly comparable to the location of reconnected field lines in the simulation, \ie\ to the current ribbons.  The close correspondence between the flare and current ribbons in the observed flare show that the sharp transition from the corona to the photosphere through the extremely complex chromospheric layer is not able to stop the coronal currents for entering into the photosphere. This is consistent with the horizontal magnetic field around the photospheric inversion line that is getting stronger during the flare, while the vertical field component is not significantly changed \citep{Sun2012,Petrie2013}.    
   
%
\section{Local increase vs global decrease of the electric currents during a flare}
\label{Sect:5}
%

%
\subsection{Evolution of electric currents, from the observations to the general understanding}
\label{Sect:5.1}

The present study of the photospheric currents during an eruptive flare event shows an increase in the current density in specific locations that define the current ribbons. The quantification of the current evolution shows in particular that the hook and the straight part of the \J-shaped current ribbons are subject to an increase during the impulsive phase of the eruptive flare. A thorough analysis of the vertical component of the currents with \Jz-maps, in relation to the vector map of the photospheric horizontal magnetic field (Section~\ref{Sect:3.2}), indicates that this increase cannot be attributed to polarization artifacts either due to accelerated particles or to anisotropic radiation.

Also, numerical MHD simulations reproducing the evolution of an eruptive flare indicate that photospheric currents further appear in locations where there was no current before (see for example the differences for the central straight part of the current ribbons in Figures \ref{figJcomparison}c,f). 

Basic theoretical arguments, however, go against these findings. Firstly, during an eruption, due to expansion, the length of the magnetic field lines increases while their end-to-end twist is conserved. This reduces (not increases) the current densities along the flux tube. Secondly, during and after the eruption, the magnetic field is classically expected to evolve toward a more and more potential state while the magnetic helicity is mostly removed by the coronal mass ejection (\ie, the flux rope) via the reconnection of coronal magnetic field lines. 

Indeed, in the 3D standard model, the total magnetic energy decreases and the initial flux rope, which is further built during the reconnection process, is ejected while simple arcades (the flare loops) are formed \citep{Aulanier2012}.  Then, the magnetic field globally relaxes to a more potential field.

{Extended currents associated with the pre-eruptive or the erupting flux rope are relatively weak in the present observation. They are actually difficult to observe: as shown in Figure~\ref{figJzzooms} \citep[and also in Figure~8 of][]{Petrie2013}, the base of the flux rope legs that are anchored in the region surrounded by the hook has very weak current densities, with a level similar to the noise level (see white and grainy regions inside the hooks in Figures \ref{figJzzooms},\ref{figBvector}). Such relatively weak current densities at the footpoints of the twisted flux rope may be explained as follows. In nearly cylindrical twisted loops, the current density is nearly inversely proportional to the structure length for a given end-to-end twist. It is therefore expected that, for a flux rope running along the PIL of an AR, the extended currents at the photospheric footprints should be relatively weaker than those of equally-sheared and shorter loops running across the PIL. These differences are well seen in numerical simulations that incorporate a long twisted flux rope and short-sheared loops (see, \textit{e.g.}, Figures 9 and 10 of \citet{Aulanier2005} for active-region loops, and Figure \ref{figJcomparison} from this paper, where extended currents at the flare loops footpoints are present between the current ribbons).}

{It is worth noticing that extended patches of strong current densities have been clearly observed from earlier ground-based magnetograms (\textit{e.g.} 
\citealt{Hagyard1988}, 
\citealt{Hofmann1988}, 
\citealt{Beaujardiere1993}). But such reports do no contradict with our present observation of barely detectable current densities at the base of the flux rope, obtained with a more recent space-borne instrument. On the one hand, old and recent observed extended current patches can simply be associated with relatively short sheared loops (as explained above). On the other hand, they may also be related with current layers associated with flare ribbons, when narrow current density patches are located along separatrices or quasi-separatrix layers footprints (see, \textit{e.g.}, \citealt{Mandrini1995} and \citealt{Demoulin1997}). The patches appear extended because of poor spatial resolution, seven times lower in these measurement compared with the HMI data. Note that the inversion technique was also improved in between, refining the measurements of the vertical current density that appear localized in the present study.}

{Interestingly, our present interpretation of the relatively weak magnitude of flux rope currents, and the difficulty of measuring them, raises questions about the possibility of recovering pre-eruptive flux ropes in non-linear force-free field extrapolations on a regular basis.
}

%
\subsection{Why do electric currents increase during a flare?}
\label{Sect:5.2}

The picture of the current evolution presented above leaves out the evolution of localized coronal current layers that are formed during the building phase, and even more during the flux rope expansion.  
In an evolving magnetic field, current sheets form on separatrices, and even more generally on QSLs \citep{Demoulin1996a}.   The formation of these currents is intrinsic to the evolution of a magnetic field with a complex topology, as the photospheric line-tying and the drastic change of field-line shape impose a different magnetic field evolution on each side of a QSL.
Then, current layers associated with separatrices/QSLs build up in eruption models from the earliest 2.5D models to the most recent 3D ones.

In models of eruptive flares, a current sheet builds up below the flux rope \citep[e.g.][and references therein]{Lin2000}.  In 2.5D, this current sheet extends around the initial magnetic X-point and stays in the corona even in a spherical geometry, because of the supposed invariance in one direction \citep[e.g.][]{Lin1998}.  However, within a 3D model, this current sheet extends all the way down to the photosphere \citep[e.g.][]{Titov1999}.  This is generalized to the build-up of a current layer at the Hyperbolic Flux Tube (HFT), which is the core of QSLs \citep{Titov02}.  

{During an evolution, slowly driven by boundary conditions, the current layer formed at the QSL becomes thinner with time.  At some point of this evolution the current layer becomes thin enough for reconnection to take place \citep{Aulanier06}.} Reconnection can take place even earlier, when the driver is an ideal instability: the evolution of the current thinning then occurs much faster, on the Alfv\'en time scale. This was shown in Figure 4 of \citet{Aulanier2012}, where the current layer below the erupting flux rope becomes rapidly thinner and reconnection occurs, forming the flare loops and further building the flux rope. 

In such a case, the driver is the torus instability, and the strength of this ideal instability implies that the coronal current layer collapses and triggers an efficient magnetic energy release via reconnection.  This also explains the impulsive nature of an eruptive flare, and the sudden increase in localized regions of the current density.


To conclude, there are two complementary, and not contradictory, evolutions of the electric currents during an eruptive flare. First, as expected by the general intuition, models of eruptive flares have shown that extended currents decrease as the magnetic configuration evolves toward a more potential state. These extended currents are less localized than the coronal current layer, and their photospheric signatures should be seen at footpoints of the flux rope magnetic field lines (inside the region surrounded by the hooks, where the flux rope legs are, see cartoon in Figure \ref{figmodel}). However, the evolution of these currents are difficult to observe because of their low intensities, but could be computed via numerical simulations.  Second, a flare also involves magnetic reconnection, taking place in current layers.  The associated flare driver further builds a thin coronal region where the current density increases. Contrary to the extended currents, these localized currents have high intensities, making them easier to observe.   The predictions from the 3D standard model of the footprints of these regions, \ie\ the current ribbons, have proven to be correct, as shown with the present analysis of the Feb. 15 eruptive flare event.
 

%
%
\section{Summary and Conclusions}
\label{Sect:6}
The present paper quantifies for the first time the photospheric current evolution during an eruptive flare and challenges the general understanding of the current density decrease associated with the evolution of the flare. Indeed, as the flare occurs, the magnetic field configuration becomes more potential, so that associated currents should globally decrease (see the discussion in Section \ref{Sect:5}). The 3D extensions of the standard eruptive flare model has successfully predicted some observational characteristics of eruptive flares, as presented in a series of papers {(see Section~\ref{sect_intro}).}
Especially, the model shows that current ribbons form at the photospheric boundaries, evolve with similar patterns to the flare ribbons, and are associated with the sudden collapse of a coronal current layer which further thinning is driven by the torus-unstable flux rope.

In this work, we aim at comparing the evolution of currents during an eruptive flare so as to confront with the results predicted with the 3D standard flare model. To do so, we have analyzed the vertical current density obtained from HMI magnetograms and with the UNNOFIT inversion code during the X-class flare event of February 15, 2011. The other components of the current density cannot be computed due to the single spectral line of the HMI observations. However, the vertical component of the current density is of highest interest, since it is used in magnetic extrapolation of coronal magnetic field, and since this component can be injected in the corona.
{Also, the 3D standard model indicates that the regions of the photospheric current ribbons are nearly force-free. The magnetic field vertical component being dominant, it results that the current density vector $\bf{J}$ is nearly $\bf{B}$-aligned, and therefore dominantly vertical too.}
The vertical current density is dominated by direct currents, especially in the flare region. The current maps and the base-difference of the direct currents are used to follow the evolution of the current ribbons. 

{The paper then reports on the following main findings for the observation of the evolution of the vertical current density:}

\begin{itemize}
\item{{During the impulsive phase, the current density increases in the hooks as well as in the straight parts of the \J-shaped current ribbons. The spatial and temporal coherence of the current ribbons shows that they are unlikely to be related to polarization artifacts. }

}
\item{{The quantification of the electric current shows increases by about 1.5 to 2 times that remains more than one hour after the occurrence of the flare peak.}
}
\item{{The structural changes of the current ribbons as seen in the observations are the same as found in the photospheric currents of the 3D MHD simulation. The two main characteristics are: first the expansion of the hook ribbon regions, that become rounder as the flux rope continues to grow, and second, the elongation of the straight part of the ribbons. }
}
\item{{The comparison of the vertical current component, in \Jz-maps, with the flare ribbons obtained from AIA 335~\AA\ filter shows similar evolutions of the current and the flare ribbons, demonstrating that {both current and flare} ribbons result from the evolution of the magnetic field.}
}
\end{itemize}

These results therefore confirm the predictions of the 3D MHD simulation of an eruptive flare and that despite not being a data driven simulation, it still incorporates at the lower boundary the key typical ingredients of an active region evolution with theoretical boundary conditions. 
In particular, it shows that the approximation of line-tied magnetic field at the photospheric boundaries is correct, as the horizontal magnetic field is still allowed to evolve.

The model allows a further understanding on the evolution of the photospheric current. A current layer forms all along the QSLs, from the corona to the photospheric layer, where the footprints can be traced as the current ribbons.  {The ideal instability of the flux rope, with the photospheric line-tying, further enhances the electric current within the current layers as it imposes a different evolution of the magnetic field on both side of QSLs.  Such fast} evolution ultimately leads to the collapse of the current layer and a fast increase of the reconnection rate.  This explains the impulsive phase of flares. 

The results also have implications on the magnetic configuration involved during an eruptive flare, as follows. The shape of flare/current ribbons is consistent with the presence of a flux rope with a twist of less, or equal to one turn. Indeed, since these ribbons follow the shape of the QSLs (see cartoon in Figure \ref{figmodel}), a more twisted flux rope structure would lead to a more complicated whirl shape for the photospheric trace of QSLs (see for example Figure 8 in \citealt{Demoulin1996b}), and therefore not to a \J-shaped structure as seen here for current ribbons ($J_z$-maps) and flare ribbons (335 \AA ). 

{The observed ribbon shapes therefore exclude the possibility to have more complicated, highly twisted, kink unstable flux ropes, as is required in some models of CME ejection (\textit{e.g.} \citealt{Gibson1998}; \citealt{Low2001} (Fig.9); \citealt{Roussev2003, Cohen2010, Kliem2010, Shiota2010, Lugaz2011, Roussev2012}). 
It should be noted that, however, those studies focus on the propagation of CMEs and not on the flare. Nevertheless, it is arguable that the development of a consistent eruption model, with the flare and the CME up to its interplanetary propagation, need to match with observed \J-shaped current and flare ribbons, since they indicate how twisted the CME magnetic field is. This is arguably important, for example because the CME acceleration depends on the Lorentz force, which itself depends on the currents, and also because the reconnection of the twisted fields with the ambient fields during the early CME expansion strongly depends on
their relative orientations.
}

The fact that the current ribbons remain coherent at the photosphere demonstrates that the chromosphere does not have a strong influence on the paths of the currents from the coronal reconnection region towards the photosphere. This was not straightforward a priori. Indeed, the increasing plasma $\beta$, through the dynamic chromosphere down to the thick photosphere, must lead to the amplification of non-negligible cross-field currents. Nevertheless, the HMI observations show that the $\beta$=0 and the infinitely thin photosphere hypotheses in the present MHD numerical simulation remain a good approximation.

{In conclusion, the electric current evolution, in the studied flare of February 15, 2011, is in agreement with the predictions made by the 3D standard model developed throughout a series of papers with the OHM code. The same model explains the evolution of flare loops \citep{Aulanier2012}, the underlying 3D reconnection process \citep{Janvier2013}, and the slipping motion of field lines during an eruptive flare event \citep{Dudik2014}. This model also quantifies the energy release in flares in function of the spatial size and magnetic flux of active regions \citep{Aulanier2013}.  Finally, all these correspondences between this 3D standard model and observations are strong evidences that the key physical ingredients are included in the model, and confirm the hypothesis of line-tied, $\beta$=0 conditions given to the model.

\begin{acknowledgements}
We thank the referee for his/her valuable comments that helped to improve the paper. AIA and HMI data are courtesy of NASA/SDO and the AIA and HMI science teams. We thank the many teams cooperative efforts for SDO/HMI and AIA. MJ acknowledges support via a contract funded by the AXA Research Fund.
This work was granted access to the HPC resources of MesoPSL financed by the Region Ile-de-France and the project Equip@Meso (reference ANR-10-EQPX-29-01) of the "Programme Investissements dÕAvenir " supervised by the Agence Nationale pour la Recherche.
\end{acknowledgements}

\bibliographystyle{apj}
\bibliography{CurrentsXflare}

\end{document}